\documentclass[aps,prl,showpacs,superscriptaddress]{revtex4}

\usepackage{graphicx}
\usepackage{amsmath,amssymb,latexsym}
\usepackage{color}
\bibstyle{apsrev.bib}

\begin{document}
\def\be{\begin{equation}}
\def\ee{\end{equation}}

\def\bfi{\begin{figure}}
\def\efi{\end{figure}}
\def\bea{\begin{eqnarray}}
\def\eea{\end{eqnarray}}

\title{On a relation between roughening and coarsening}

\author{Federico Corberi}
\affiliation {Dipartimento di Fisica ``E.~R. Caianiello'', and INFN, Gruppo Collegato di Salerno, and CNISM, Unit\`a di Salerno,Universit\`a  di Salerno, 
via Giovanni Paolo II 132, 84084 Fisciano (SA), Italy.}

\author{Eugenio Lippiello}
\affiliation{Dipartimento di Scienze Ambientali, Seconda Universit\`a di Napoli,
Via Vivaldi, Caserta, Italy.}

\author{Marco Zannetti}
\affiliation {Dipartimento di Fisica ``E.~R. Caianiello'',Universit\`a  di Salerno, via Giovanni Paolo II 132, 84084 Fisciano (SA), Italy.}

\date{\today}

\begin{abstract}
We  argue that a strict relation exists between two in principle unrelated quantities:
The size of the growing domains in a coarsening system, and the kinetic roughening of
an interface. This relation is confirmed by extensive simulations of the Ising model with
different forms of quenched disorder, such as random bonds, random fields and
stochastic dilution.
\end{abstract}

\pacs{05.40.-a}

\maketitle

Slow relaxation is a 
feature found in a variety 
of physical systems including  randomly stirred fluids, ballistic
aggregation, magnetic flux lines in superconductors \cite{sug_pl1},  
directed polymers and manifolds in random 
media \cite{Kard97,Kard87,Fish91},  
phase-ordering 
\cite{puri09,Puri04,Bray91,corbcrp,zanreview,Paul04,Henk08,Sic08,Corb02,Lip10,Corb11,Corb12,Corb13,sug_pl2}
and others.
In these cases
the evolution is characterized by scale-free 
power-law behaviors and the equilibrium state is approached on 
a timescale that diverges  in the thermodynamic limit. 
In this Letter we focus on two paradigms of slow relaxation,
i) the evolution a disordered ferromagnet
after a temperature quench and ii)
the roughening of an interface in a medium.
The aim of the paper is to show
a strict correspondence between i) and ii).

A unified description of the two problems above can be arrived at by
considering a ferromagnetic system, namely 
a collection of interacting Ising spins $S_i$ on the sites $i$ of a lattice.
For such a system the usual
para-ferromagnetic transition occurs at some critical temperature $T_c$.
Quenched disorder, in the form of random fields or bonds, stochastic dilution or
other sources of randomness, can be present provided that the low-temperature ordered phase 
is preserved.
The two kinds of slow evolution i) and ii) mentioned above 
can be observed in such a model at low temperature $T<T_c$
by preparing the sample with the initial condition (at time $t=0$) 
in the following two ways:

i) The value of  each spin $S_i=\pm 1$ is randomly chosen 
and is uncorrelated from the others, corresponding to the 
equilibrium state of the ferromagnet at infinite temperature. 
This protocol amounts to the 
instantaneous quench of the system from the initial temperature $T_i=\infty$ 
to the final temperature $T<T_c$. As it is well known, a coarsening stage is observed 
\cite{puri09,Puri04,Bray94,Bray91,corbcrp,zanreview,Paul04,Henk08,Sic08,Corb02,Lip10,Corb11,
Corb12,Corb13,sug_pl2} with 
growing magnetic domains of size ${\cal L}(t,\epsilon,\ell)$, where
$\ell $ is the system size and
we indicate generically by $\epsilon $ the strength of the disorder.
For example, denoting 
by $L$ the domains size in an infinite system 
($L(t,\epsilon)\equiv {\cal L}(t,\epsilon,\ell=\infty)$),
in a pure (non disordered) magnet ($\epsilon =0$) one usually has 
$L(t,0)\propto t^{1/z}$ (with $z=2$
with a non-conserved order parameter).
Dynamical scaling \cite{Bray94} implies that $L$ is the only dominant length
at large times.

ii) The system is divided into two halves, as for instance by the diagonal in a 
square  system, and spins are set to the value $S_i=+1$ in one half and 
$S_i=-1$ in the other. Appropriate boundary conditions are provided in such a 
way that the spanning interface seeded by the initial condition remains at all times
(e.g. if the system is divided by the diagonal anti-periodic boundary conditions
are used).
The main feature of the process ii) is the 
kinetic roughening of the interface which, in a pure system, 
is described by the usual Family-Vicksek scaling relation \cite{FV,Kolt}
\be
W(t,\ell)=A(T)\,\xi  (t)^{\alpha} {\cal V}\left (\frac{ \ell }{\xi (t)}\right ),
\label{central}
\ee
where $W(t,\ell)$ is the interface width, or roughness, namely the root mean 
square displacement with respect to the initial straight configuration
(see further on for an operative definition),
$\xi (t)=t^{1/Z}$ is a quantity with dimensions of a length, 
$A$ is a temperature-dependent constant 
(the width of an interface in equilibrium in a box
of unitary length $\ell =1$), $\alpha $ and $Z$ are the roughening and the dynamical exponent. 
For the Edwards-Wilkinson 
universality class, corresponding to an interface  
in a magnetic system without conservation of the order parameter, one has
$Z=2$.
${\cal V}(x)$ is a scaling function with the limiting behaviors
\be
\left \{
\begin{array}{l}
{\cal V}(x)= const.\hspace{.65cm},\hspace{.5cm} \mbox{for}\,\,\, x\gg 1\\
{\cal V}(x)= x^\alpha \hspace{1.2cm},\hspace{.5cm} \mbox{for}\,\,\, x\ll 1
\end{array}
\right .
\label{asbev}
\ee    

Let us remark that $W(t,\ell)$, considered in the evolution i) 
must not be confused with the width of interfaces separating different domains
during the coarsening evolution i). However, both   
the evolutions i) and ii) are determined by the flip of spin on interfaces
which are governed by the same microscopic rules. This will be important in the following.
Notice also that in the pure case, by running the two processes i) and ii) from $t=0$ in parallel 
and measuring $L$ in i) and $\xi$ in ii) one would find
\be
L(t,\epsilon=0)\propto \xi(t)
\label{lispropxi}
\ee
because $L\sim t^{1/z}$ and $\xi \sim t^{1/Z}$ with $z=Z$ in this case.
This result highlights the robustness of the fundamental
length responsible of dynamical scaling under strong variations of the
initial condition, as when passing from system i) to ii). In this paper we shall carry
out a detailed study of this remarkable result by extending the
investigation to the vast area of systems with quenched
randomness.

An important point that we want to address is how Eqs. (\ref{central},\ref{asbev}) change
in the presence of quenched randomness.
Quite generically \cite{Lip10,Corb11,Corb12,Corb13,Corb02,corbcrp} this additional source of disorder
introduces an extra length $\lambda (\epsilon )$ in the problem. The actual meaning of this quantity
depends on the system at hand: In the presence of quenched impurities, for instance, $\lambda $ can be
associated to the typical distance among two of them. 
One has usually the property
\be 
\lim _{\epsilon \to 0} \lambda (\epsilon)=\infty,
\label{property}
\ee 
as it can be easily understood in the case of quenched impurities.
Clearly, a possible generalization of Eq. (\ref{asbev}) must keep into account the
presence of this additional length $\lambda $. Furthermore, it is well known \cite{schehr} that disorder modifies the growth of correlations $\xi$
(as will be clear from our data, see discussion later)
with respect to the behavior in a pure system.

\begin{figure}[h]
\vspace{2cm}
\begin{center}
\includegraphics[width=1\columnwidth]{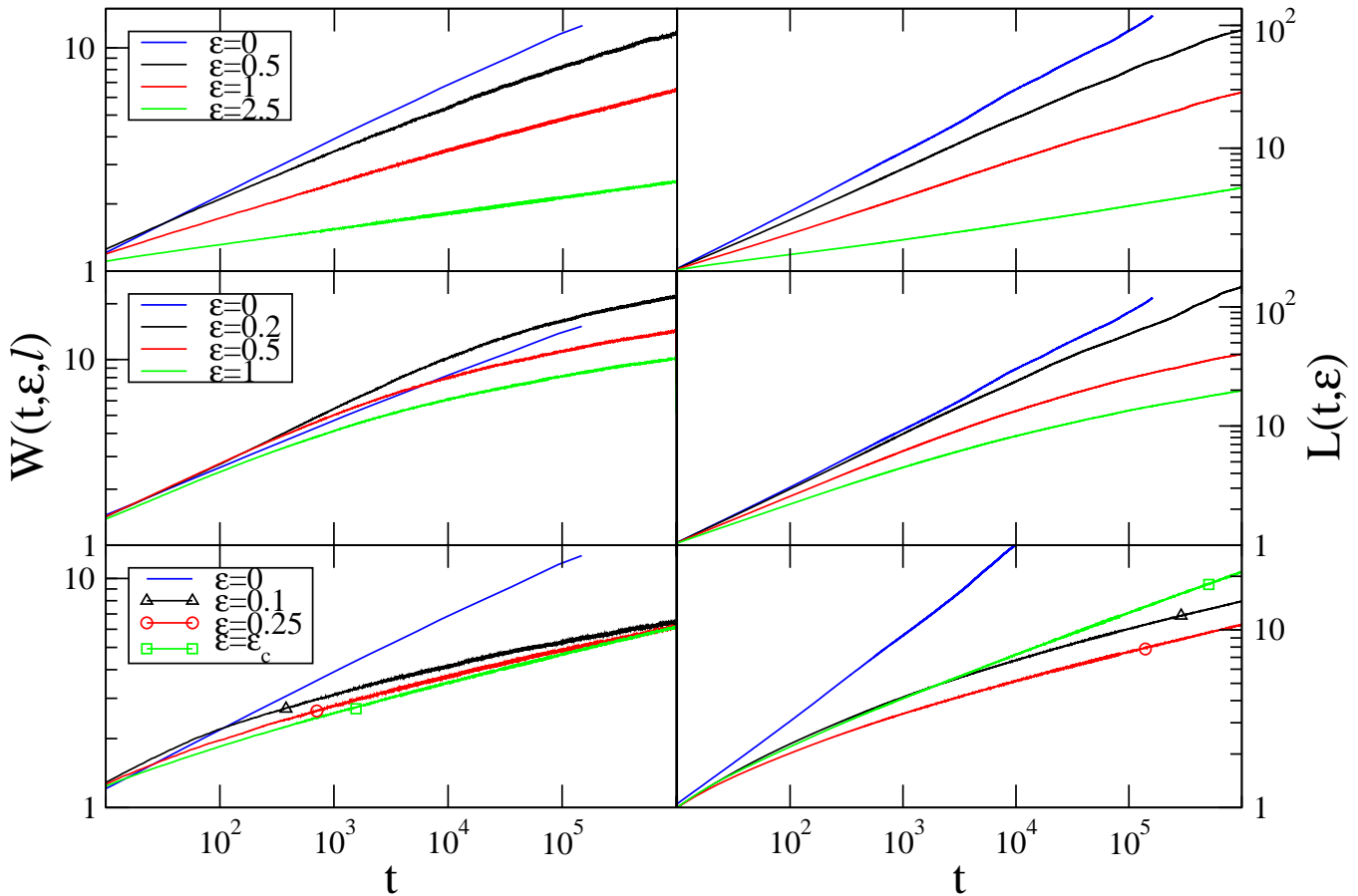}
\end{center}
\caption{
(Color online) $W(t,\epsilon,\ell=\infty)$ (left three panels) and $L(t,\epsilon)$ 
(right three panels) are plotted against time on a double-logarithmic plot
for the RBIM (top two panels), the RFIM (central two panels) 
and the SDIM (bottom two panels).
For the RBIM the values of disorder displayed are $\epsilon=0$ (pure case, blue), $\epsilon=0.5$ (black),
$\epsilon=1$ (red), $\epsilon=2.5$ (green), from top to bottom.  
For the RFIM the disorder values are $\epsilon=0$ (pure case, blue), $\epsilon=0.2$ (black),
$\epsilon=0.5$ (red), $\epsilon=1$ (green), from top to bottom. 
For the SDIM the dilution values are $\epsilon=0$ (pure case, blue) 
$\epsilon=0.1$ (black with a triangle), $\epsilon=0.25$ (red with a circle) and $\epsilon=\epsilon_c$ 
(green with a square).}
\label{fig_wrong}
\end{figure}
Borrowing from the fact that in the pure case one can replace $\xi $
with the domains size $L $ of the corresponding process i), according to Eq. (\ref{lispropxi}),
we infer that the proper generalization of the scaling (\ref{central}) to systems with
quenched disorder is a two-parameter scaling with typical lengths $L(t,\epsilon)$
and $\lambda (\epsilon)$, namely
\be
W(t,\epsilon,\ell)=A(T,\epsilon)\, L (t,\epsilon)^\alpha  
{\cal W}\left (\frac{ \ell }{L (t,\epsilon)},\frac{\lambda (\epsilon)}
{L(t,\epsilon)} \right ),
\label{wvsld}
\ee
with a scaling function such that 
\be
\lim _{y\to \infty}{\cal W}(x,y)={\cal V}(x),
\label{bescf}
\ee
where ${\cal V}(x)$ is given in Eq. (\ref{asbev}) in order to recover Eq. (\ref{central})
when disorder is absent (using Eq.  (\ref{property})). 

Eq. (\ref{wvsld}) is a bridge between two {\it a priori} unrelated quantities, the width of 
an isolated roughening interface (on the l.h.s.) and the size of the domains in a 
coarsening process (on the r.h.s). 
It is the central statement of this paper: Its verification 
in different systems will be the main issue below. 
Before doing that, let us
stress that the form (\ref{wvsld}) is different from a usual two-parameter scaling,
because the strength of disorder 
enters not only through the extra length 
$\lambda (\epsilon)$ but also in the quantity $L(t,\epsilon)$.  
We will show that
Eq. (\ref{wvsld}) describes the whole pattern of behaviors of magnetic systems with different 
form of quenched disorder and for any strength $\epsilon $ of their randomness.

To start with, Eq. (\ref{wvsld}) is obviously correct in the  pure case $\epsilon =0$, 
since due to Eqs. (\ref{property},\ref{bescf})
one recovers Eq. (\ref{central}) when disorder is absent. This
shows that, both in Eqs. (\ref{central}) and (\ref{wvsld}), $\alpha $ is
the roughening exponent of the pure case which is obviously independent
of disorder.
In the opposite limit $\epsilon \to \infty$, when disorder is very strong, the roughening 
exponent is expected to take a different value $\zeta \neq \alpha$ \cite{Kard97}. 
This is accounted for by the scaling (\ref{wvsld}) if 
$\lim _{y\to 0}y^{\zeta - \alpha}{\cal W}(x,y)\sim \omega (x)$, where $\omega $ is 
another function with asymptotic behaviors analogous to those (\ref{asbev}) of
${\cal V}$. Summarizing, one has
\be
W(t ,\epsilon,\ell)=A(T,\epsilon)\cdot \left \{
\begin{array}{l}
L (t,\epsilon)^\alpha \, {\cal V}(x)\,,\hspace{1.2cm}\mbox{for}\,\, y\gg 1 
\,[w]\\
L (t,\epsilon)^\zeta \,\lambda (\epsilon) ^{\alpha -\zeta}\, \omega(x)\,,\mbox{for}\,\, y\ll 1 
\,\, [s],
\end{array}
\right .
\label{asbev1}
\ee 
where the two regimes $y\gg 1$ and $y\ll1$ are denoted as $w$ and $s$ for the reasons that will be 
explained below.

In the following we will check the validity of Eq. (\ref{wvsld}) in 
the two-dimensional Ising model with
random fields, random bonds and site dilution.
All these models can be described by the
Hamiltonian $H=-\sum _{\langle ij\rangle}J_{ij}n_{i} n_j S_i S_j -\sum _i h_i S_i$ 
where $i$ ($j$) are sites on a square lattice and $\langle ij\rangle$ are
nearest neighbors couples.
From this Hamiltonian the random bond Ising model (RBIM) \cite{corbcrp,Corb11} 
is obtained by taking $n_i=1$ $\forall i$, $h_i=0$ $\forall i$ and $J_{ij}$ random 
variables uniformly distributed 
in $[J-\epsilon \,T, J+\epsilon \,T]$, with $\epsilon \,T<J$. 
The random field Ising model (RFIM), on the other hand, amounts to
$n_i=1$ $\forall i$, $J_{ij}=J$ $\forall \langle ij\rangle$ and $h_i=\pm \epsilon \,T$ 
uncorrelated random variables with equal probability. Finally, the site 
diluted Ising model (SDIM) is obtained by taking a uniform $J_{ij}=J$, 
$h_i\equiv 0$ and $n_i$ uncorrelated random variables such that
$n_i=0$ with probability $\epsilon $ (the dilution) and $n_i=1$ with probability $1-\epsilon $. 

The systems above are evolved numerically, both for process i) and ii) by means of single spin 
flips with 
Glauber transition rates up to $10^6$ Montecarlo steps.
Besides,
we prevent the flipping of bulk spins
which are aligned with all the nearest neighbors.
In case ii) this guarantees that the interface is unique at all times.
In case i) this modified dynamics has been
used in several studies \cite{nobulk} of phase-ordering where 
it was shown that it does not
alter the behavior of the quantities we are interested in,
in the limit of small temperatures that will be considered below. 
It has also the advantage to speed-up the computation. 

Let us start by discussing the details of the simulations for the phase-ordering kinetics,
namely process i).
For this problem we have performed a set of numerical runs where phase ordering dynamics 
occurs in a system of size $\ell=2000 \,a_0$, where $a_0$ is the lattice-spacing and we
set $a_0=1$ hereafter. 
Since we have checked that this sample is sufficiently large to avoid finite-size effects in the time range considered, one has
${\cal L}(t,\epsilon,\ell =2000) \simeq {\cal L}(t,\epsilon,\ell =\infty) = L(t,\epsilon)$.
The characteristic size $L $ -- one building block of Eq. (\ref{wvsld}) -- is obtained 
as the inverse excess energy
$L (t,\epsilon)\sim [E(t)-E(\infty)]^{-1}$,
where $E(\infty)$ is the energy of the target equilibrium state 
\cite{Lip10,Corb11,Corb13} and $E(t)$ is the energy at time $t$.
This is a standard way to evaluate $L (t,\epsilon)$ in coarsening systems: it is based on the
fact that the interior of domains is in equilibrium and  
the excess energy is stored on the
interface, so $E(t)- E(\infty)$ is proportional to their total length.
This in turn is given by the length of a single domain's boundary ($\propto L$)
times the number of such domains ($\propto L^{-2}$), from which the above
relation between $L (t,\epsilon)$ and the excess energy is obtained.

Phase-ordering in random ferromagnets has been the subject of a number of
recent studies \cite{Paul04,Henk08,Sic08,Corb02,Lip10,Corb11,Corb12,Corb13,corbcrp,sug_pl2} to 
which we refer for 
a detailed discussion. For the scope of the present study what has to be
recalled is the existence of a crossover length $\lambda (\epsilon)$ 
such that $L (t,\epsilon)$ crosses over from an early regime where 
the growth-law is still algebraic (although with an exponent that may depend
on $\epsilon$ and $T$) for $L (t,\epsilon)\ll \lambda (\epsilon)$,
to a late logarithmic growth $L (t,\epsilon)\gg \lambda (\epsilon)$. 
We will refer to these two regimes in a compact way as $w$ 
(standing for {\it weak}, as referred to 
the effects of disorder, since the growth-law remains algebraic) and 
$s$ (standing for {\it strong}), respectively. Let us also mention that
$T_c=0$ for the RFIM in $d=2$, meaning that coarsening in this system 
is interrupted after a time which diverges in the $T\to 0$ limit.
In all the cases considered here this time is far beyond the times accessed in the
simulations.

Regarding process ii), indicating with $(m,n)$ the
(horizontal and vertical) coordinate of a lattice site $i$,
the width is evaluated as $W(t,\epsilon,\ell)=(\ell/a_0)^{-1} \sum _{n=1}^{\ell} \sqrt {w_n(t)^2} $.
Here $ w_n(t)$ is the distance travelled by the portion of interface, with vertical coordinate  $n$, in the time interval $[0,t]$.

In order to check the scaling 
relation (\ref{wvsld}), for the single interface, we have considered different system 
sizes $\ell \in [4 ,1000]$. We observe that, in the $T\to 0$ limit we consider,
the interface roughens almost independently of $T$. This is at variance with what observed
in continuum theories of kinetic roughening \cite{upton}, meaning that
the discrete character of our model is relevant at such low temperatures.

Due to the presence of the two arguments in the scaling function of 
Eq. (\ref{wvsld}) its verification would in principle require, for any of the
disordered models introduced above, a complete scan
of the three-dimensional parameter-space $t,\epsilon,\ell$ and the determination
of the scaling arguments $x=\ell /L$, $y=\lambda /L$.
Since this is out of reach, we mainly focus on the limits $y\gg 1$ and $y\ll 1$,
corresponding to the regimes $w$ and $s$ defined above, so that Eq. (\ref{wvsld})
simplifies to the couple of forms in (\ref{asbev1}) where the scaling functions
${\cal V}$ and $\omega$ depend on a single scaling variable. 

{\bf RBIM)} We start this program from the RBIM. In this case it was shown in~\cite{Corb11}
that the crossover from the regime $w$ to the one $s$ is delayed to such huge times that in simulations 
the condition $\lambda (\epsilon)/L (t,\epsilon)\gg 1$ is always met, 
namely the system is described at any reasonably accessible time by 
the regime $w$ of Eq.~(\ref{asbev1}).
This is very neatly observed in the upper-right panel of Fig. \ref{fig_wrong}, where
power growth-laws of $L(t)$ are observed in the range of simulated times and there is
no indication of a crossover to a slower (i.e. logarithmic) growth, typical of
regime $s$, for any
choice of $\epsilon $.
Therefore, by plotting $L (t,\epsilon)^{-\alpha}\,W(t,\epsilon,\ell)$,
with the $d=2$ Edwards-Wilkinson roughening exponent $\alpha =1/2$, 
against $x=\ell /L (t,\epsilon)$ for any choice of $\epsilon$ one should 
find data collapse among curves obtained with different values of $\ell$.
Furthermore, upon changing $\epsilon$ one expects only to observe a different 
prefactor $A(\epsilon, T)$ which, plotting on a double log scale, amounts
to a vertical shift. 

\begin{figure}[h]
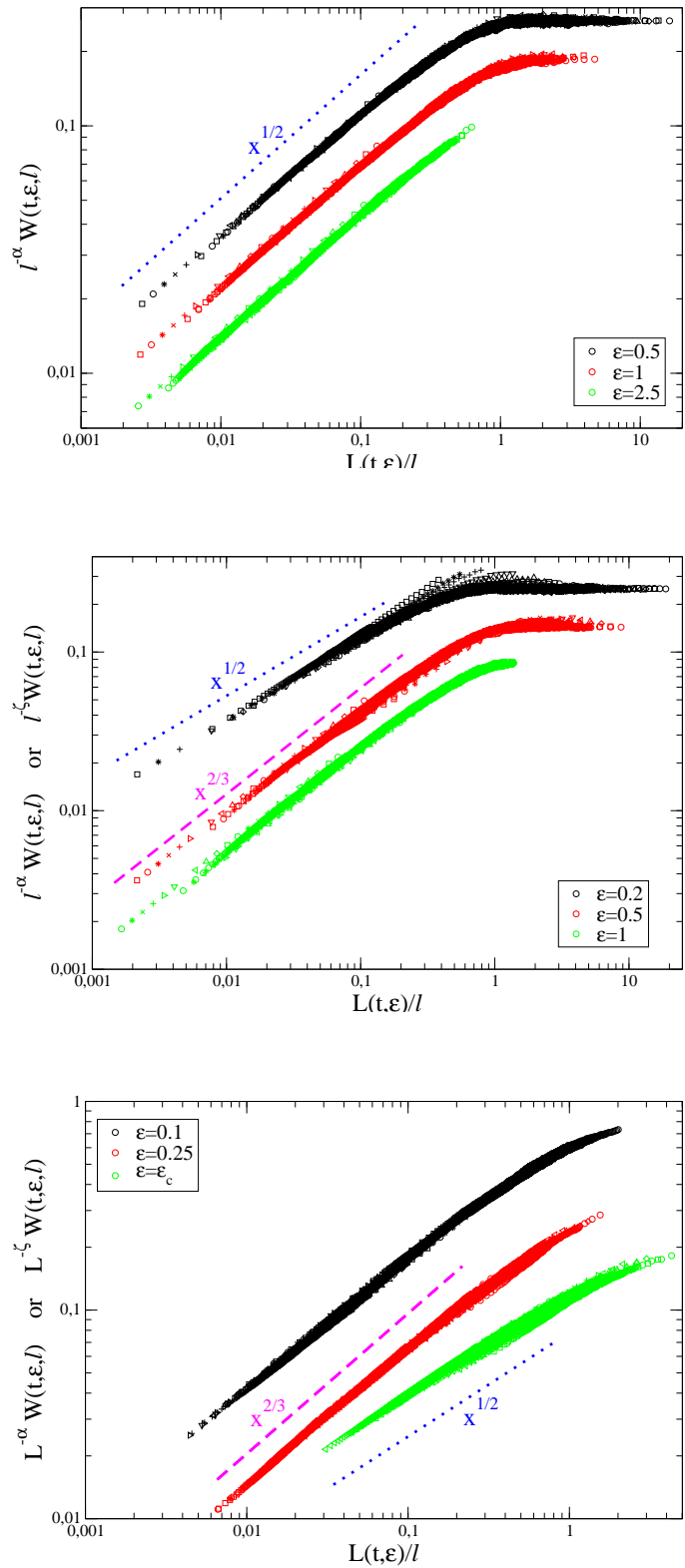

\begin{center}
\includegraphics[width=.5\columnwidth]{rbim_scal96.eps}

\vspace{1cm}

\includegraphics[width=.5\columnwidth]{rfim_scal96.eps}

\vspace{1cm}

\includegraphics[width=.5\columnwidth]{sdim_scal96.eps}
\end{center}
\caption{
(Color online) Upper panel- 
The parametric plot $\ell ^{-\alpha} W(t,\epsilon,\ell)$
vs  $L(t,\epsilon)/\ell$ is plotted for the RBIM with three disorder values $\epsilon=0.5$ (black), $\epsilon=1$ (red), $\epsilon=2.5$ (green) from top to bottom.  For each $\epsilon$ we consider different values of $\ell$ given by $\ell=5+1.2^j$, where $j$ is an integer in the range $[0:23]$,  
 represented by different symbols.
The blue-dotted line is the power law behavior $x^{1/2}$.
Here and in the other panels curves relative to different values of $\epsilon $ are 
displaced vertically (data are multiplied by an arbitrary constant) for a better
presentation. 
Central Panel - As in the upper panel for the RFIM with disorder values $\epsilon=0.2$ (black),
$\epsilon=0.5$ (red), $\epsilon=1$ (green) from top to bottom. 
For this model, data collapse is obtained by plotting $\ell ^{-\alpha}W(t,\epsilon)$ 
for $\epsilon=0.2$ and $\ell ^{-\zeta}W(t,\epsilon)$ for $\epsilon=0.5$ and $\epsilon=1$. 
The blue-dotted line and the magenta dashed lines are the power law behavior $x^{1/2}$ 
and $x^{2/3}$, respectively. 
Lower Panel - As in the other panels but for the SDIM for three different dilution values 
$\epsilon=0.1$ (black), $\epsilon=0.25$ (red) and $\epsilon=\epsilon_c$ (green).
For this model, data collapse is obtained by plotting $\ell ^{-\alpha}W(t,\epsilon)$ 
for $\epsilon =\epsilon _c$ and $\ell ^{-\zeta}W(t,\epsilon)$ for 
$\epsilon=0.1$ and $\epsilon=0.25$. 
The blue-dotted line and the magenta dashed lines are the power law 
behaviors $x^{1/2}$ and $x^{2/3}$, respectively. }

\label{fig3}
\end{figure}

As shown in the first panel of Fig. \ref{fig3} 
all these features are very well verified by the numerical data
(notice that we use here $1/x$ on the horizontal axis, not $x$). 
Moreover, the exponent 
$\alpha =1/2$ can also be read out from the large $x$ behavior 
(left part of the figure) of the 
scaling function, which grows as $x^\alpha$ according to Eq. (\ref{wvsld}) and
to the second of Eqs. (\ref{asbev}) (this 
is drawn as a blue-dotted line in the figure). 
We stress the highly non-trivial nature of the scaling we have verified, 
since it allows us to collapse (apart from the shift due to $A$) curves 
relative to different disorder strengths for which the growth-exponent
of the $L$'s (and hence of $W$)
varies by a factor as large as three, as can be detected in the upper-right panel of Fig. \ref{fig_wrong}. 

Regarding the constant $A$ it turns out to be very weakly dependent
on $\epsilon$. Indeed the curves for different values of $\epsilon $ are basically 
superimposing (however, for a better presentation, in Fig. \ref{fig3} curves with different $\epsilon $ have
been displaced vertically, see caption). A similar behavior of
$A$ is found also for the other two models (RFIM and SDIM)
discussed below.

Let us now comment on what we anticipated below Eq. (\ref{property}),
namely the fact that in system ii) the growth of  correlations in the presence of disorder
is modified with respect to the behavior $\xi (t)$ holding in the pure case. 
Indeed, if the growth was the same as in the pure case,
and since $\alpha=1/2$, one should observe an increase 
$W(t,\epsilon,\ell=\infty)\sim \xi (t)^\alpha \sim t^{1/4}$ of the roughness in an infinite system
in the regime $w$ we are considering, according to Eqs. (\ref{property},\ref{bescf}).
This is clearly observed in the upper-left panel of Fig. \ref{fig_wrong} only in the pure case
$\epsilon =0$, as expected, while a completely different behavior is displayed for any disorder strength 
$\epsilon \neq 0$. One arrives at the same conclusion
by considering the different disordered 
models that are discussed below.

{\bf RFIM)} Due to its delayed crossover the RBIM was previously 
used as a benchmark to verify 
Eq.~(\ref{asbev1}) in the regime $w$. In order to access the regime $s$,
we will now focus on the RFIM where, since~\cite{Corb12} 
$\lambda (\epsilon) \simeq \epsilon ^{-2}$, one can arrange the parameter
$\epsilon$ (the amplitude of the random field (in units of $T$), as specified below
Eq. (\ref{asbev1})) 
in order for the crossover from $w$ to $s$ to occur in the range of simulated times.
Before doing that, in order to match with what we observed for the RBIM, we set
$\epsilon $ to a value as small as $\epsilon =0.2$ for which 
the crossover time is larger than the simulated times. This is confirmed by
inspection of the central-right panel of Fig. \ref{fig_wrong}). Here one sees
that also for such a small value of $\epsilon $ the behavior
of the system is radically different from the non-disordered case, since $L$ grows
with an effective exponent $1/z\simeq 0.4$, definitively smaller than for $\epsilon=0$.
However, there is no manifestation of a crossover to the slower (i.e. logarithmic) growth-law
characteristic of the $s$ regime (which,
instead, is clearly observed in the same panel for the two larger values of $\epsilon$, as
will be discussed later).
In this small-$\epsilon$ case, since we are yet confined into the $w$ regime,  
we expect to observe a situation analogous to one discussed
previously regarding the RBIM. This is in fact what one observes in 
the second panel of figure~\ref{fig3} (upper set of data, in black): curves for different values of
$\ell $ can be collapsed by using the rescaling $w$ of Eq.~(\ref{asbev1}),
with $\alpha =1/2$. We interpret the somewhat poorer quality of the
collapse, as compared to the RBIM, as due to the fact that even for such
a small value of $\epsilon $ we have not been able to push the crossover
time well beyond the end of the simulation. 

The situation is radically different
when values of $\epsilon $ as large as $\epsilon =0.5$ and $\epsilon =0.8$
are used. In this case the central-left panel of Fig. \ref{fig_wrong} shows
that a crossover from $w$ to $s$ is observed quite early.
Actually, it occurs so early that collapsing the
curves by means of the scaling $w$ of Eq. (\ref{asbev1}) turns out to be
impossible. Instead a good mastercurve is obtained by using the form $s$
of Eq.(\ref{asbev1}), with $\zeta =2/3$.
We recall that this is the asymptotic roughening exponent expected for a class of
models where disorder acts only on the interfaces \cite{noirough,surfbulk}.
The same exponent is also observed pre-asymptotically in the present model \cite{noirough}, despite
an asymptotic value $\zeta =1$ is expected for very large times \cite{noirough,surfbulk}. 
The good scaling obtained with $\zeta =2/3$ signals that the truly asymptotic 
behavior with $\zeta =1$ is located much further in time.
Again, we stress the notable character of the collapse as compared to
the very different forms of $L$ and of $W$ shown in the central panels
of Fig. \ref{fig_wrong}, as $\epsilon $ is changed. 

{\bf SDIM)} 
It is known that 
in this model disorder introduces two characteristic 
lengths \cite{Corb13}. 
In a nutshell, $\lambda _1 (\epsilon) \propto \epsilon ^{-1/d}$ 
represents the average distance between vacancies, while 
$\lambda _2 (\epsilon)\propto (\epsilon _c -\epsilon)^{-\nu}$,
where $\epsilon _c=1-p_c\simeq 0.4072$ is the vacancy density at percolation
and $\nu = 4/3$ the critical exponent, represents the coherence length 
over which the fractal percolative structure extends. 
In the presence of two lengths associated to the disorder Eq. (\ref{wvsld})
can be generalized to
\be
W(t,\epsilon,\ell)=A(T,\epsilon)\, L (t,\epsilon)^\alpha  
{\cal W}\left (\frac{ \ell }{L (t,\epsilon)},\frac{\lambda_1 (\epsilon)}
{L(t,\epsilon)},\frac{\lambda_2 (\epsilon)}
{L(t,\epsilon)}\right ).
\label{wvsld2}
\ee  
The SDIM offers an interesting example where disorder introduces
a couple of lengths
allowing one to verify Eq.~(\ref{wvsld2}) in this more complex framework.

As it is clear from their
behavior, in the opposite limits $\epsilon \to 0$ or $\epsilon \to \epsilon _c$
either one or the other length diverges, meaning that -- when this occurs -- the regime $w$ of 
Eq.~(\ref{asbev1}) extends to very long times. On the other hand, for 
values of $\epsilon $ well inside the range $[0,\epsilon _c]$, such as,
for instance, $\epsilon =0.25$, any length associated to the disorder
becomes small and the regime $s$ is entered very soon. 
A first consequence of this structure is that for $\epsilon =0.25$ the growth-law
of $L(t,\epsilon)$ turns out to be slower, since it is soon logarithmic, than
in the cases closer either to $\epsilon =0$ or $\epsilon =\epsilon _c$. 
This is very clearly seen in the lower-right panel of 
Fig. \ref{fig_wrong}.
A second consequence is that we expect to observe the form $w$ of Eq.~(\ref{asbev1})
for $\epsilon =0$ or $\epsilon _c - \epsilon \ll 1$
(say, $\epsilon =\epsilon _c$), while for intermediate values of $\epsilon $ collapse
should be obtained with the scaling $s$ of Eq.~(\ref{asbev1}).
This is very well observed in the third panel of Fig.~\ref{fig3},
confirming again Eq.~(\ref{wvsld2}).  
Let us emphasize that this result,
obtained in a case with a 
complex structure with more than one length associated to
quenched disorder, is a highly non-trivial and stringent check 
providing a further 
strong indication of the general validity of Eq.~(\ref{wvsld2}). 

\begin{figure}[h]
\begin{center}
\includegraphics[width=0.3\columnwidth]{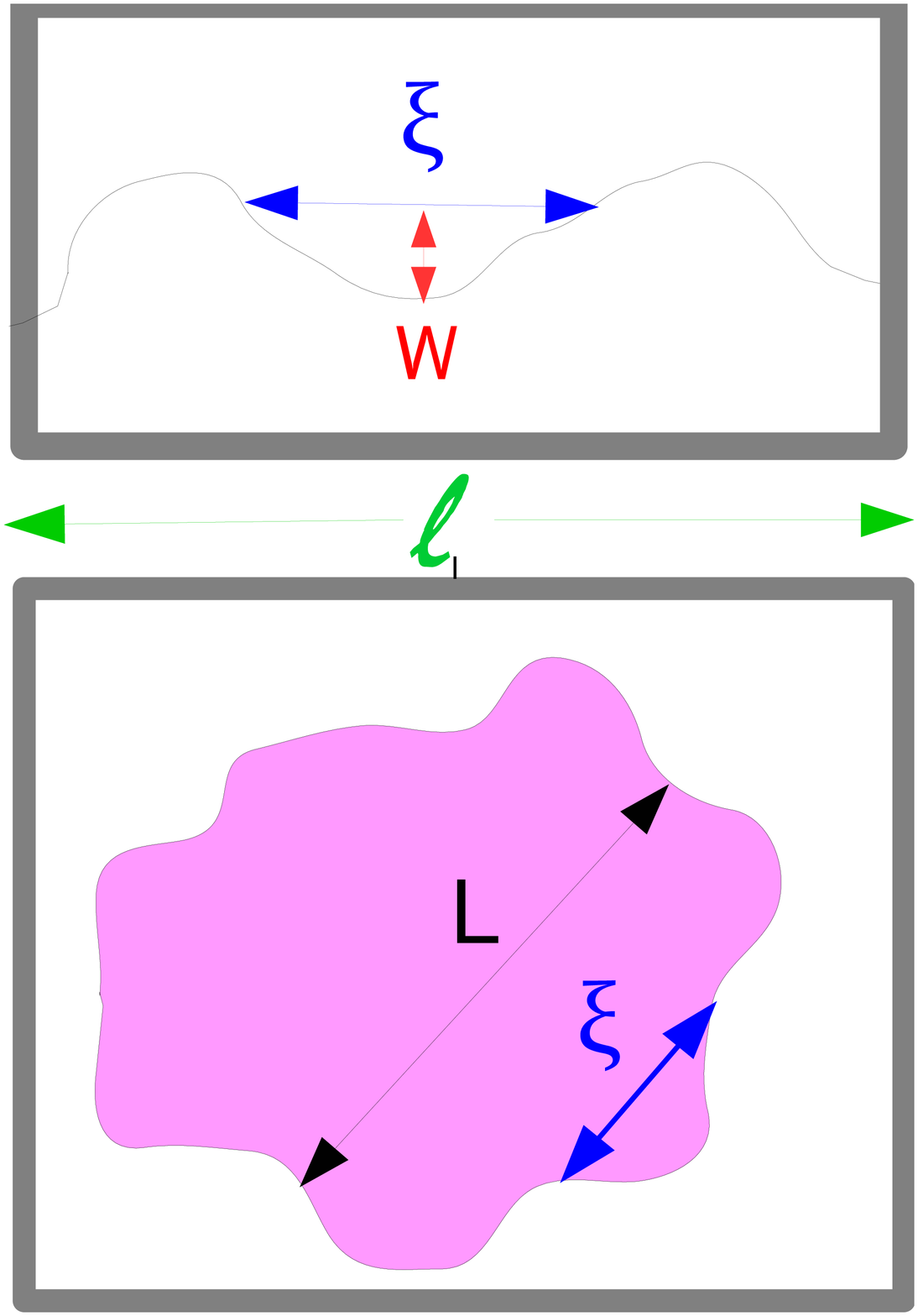}
\end{center}
\caption{
(Color online) Sketch of a single interface spanning a system (upper part) and of a domain during coarsening (lower part).} 
\label{domain}
\end{figure}

Eq. (\ref{wvsld}) establishes a direct correspondence between roughness of interfaces and 
size of growing domains holding in a variety of systems, pure or with different kind
of quenched randomness. One might wonder which is the physical reason behind this.
A possible explanation is the following: focusing for simplicity on the pure case, 
it is well known that the quantity $\xi (t,\epsilon =0)$  appearing in Eq. (\ref{central}) 
describes a longitudinal (i.e. along the surface) coherence length growing
indefinitely in a system of infinite size ($\ell=\infty$). 
Since the two processes i) and ii) are governed by a unique underlying model, 
an analogous coherence length could be expected to develop also on the 
boundaries of  the coarsening domains of case i) (still with $\ell =\infty$), 
as shown pictorially in the lower part of Fig. \ref{domain}. Dynamical scaling \cite{Bray94}
implies that there must be a unique relevant length, suggesting
$\xi \simeq L$. 
This shows that $\xi $ and $L$ are basically the same object and 
that their common behavior $\xi \sim L \sim t^{1/z}$ in the pure case is not a coincidence,
allowing one to interchange them when writing the Family-Vicksek scaling (\ref{central}).
A similar argument to infer the roughness of domain's boundaries in a coarsening system
is quoted in \cite{upton}. This basically assumes that each interface in system i) can be mapped 
on one of those of process ii) in equilibrium in a system whose size $l$ is kept adiabatically equal
to the domain's size $L(t)$ of process i). Since in equilibrium it is $W^{eq}(l)\sim l^\alpha$,
this implies that in a coarsening system
the domain walls have a roughness
$W(t)\sim W^{eq}(l=L(t))\sim L^\alpha (t)\sim t^{\alpha/z}$. Our statement is partly different.
Indeed the argument we propose only relies on the presence of dynamical scaling, and there is no
need to make any equilibration hypotheses. Furthermore our conclusions are more general,
since on one hand Eq. (\ref{wvsld}) applies not only to the interfaces of a coarsening system,
but also to different situations as the one encountered in the process ii) and others. 
Furthermore Eq. (\ref{wvsld}) is a complete scaling form describing the process from the 
out-of-equilibrium stage up to the stationary asymptotic state of the interface as well as the 
intermediate crossover, 
incorporating in addition also the finite size effects due to
$\ell <\infty$. 
Clearly, the argument above -- repeated in the presence of quenched randomness with the further ingredient
of the extra length $\lambda$ -- generalizes the conclusions above to disordered systems.

In this paper we have shown a strict relation between two in principle unrelated quantities,
the size of the growing domains in a coarsening system, and the roughening of
an interface. This relation, which has been confirmed numerically 
in a variety of disordered systems,
opens the way to a cross-fertilization between the fields of phase-ordering and
kinetic roughening, allowing one to borrow knowledge from one side to the other, possibly
improving the state of the art of our comprehension of these processes.
Because of that, a deeper understanding of this relation and of its domain of applicability 
would be desirable and represents an interesting
subject for future research.

\end{document}